# A mapping-free NLP-based technique for sequence search in Nanopore long-reads

Tomasz Strzoda[1], Lourdes Cruz-Garcia[2], Mustafa Najim[2], Christophe Badie[2] and Joanna Polanska[1,*]

[1]Department of Data Science and Engineering, Silesian University of Technology, Gliwice, Poland, [2]Cancer Mechanisms and Biomarkers Group, Centre for Radiation, Chemical and Environmental Hazards, UK Health Security Agency, Oxfordshire, OX11 0RQ

*To whom correspondence should be addressed.

**Contact:** joanna.polanska@polsl.pl

## Abstract

**Background:** In unforeseen situations, such as nuclear power plant's or civilian radiation accidents, there is a need for effective and computationally inexpensive methods to determine the expression level of a selected gene panel, allowing for rough dose estimates in thousands of donors. The new generation in-situ mapper, fast and of low energy consumption, working at the level of single nanopore output, is in demand. We aim to create a sequence identification tool that utilizes Natural Language Processing (NLP) techniques and ensures a high level of negative predictive value (NPV) compared to the classical approach.

**Results:** The training dataset consisted of RNASeq data from 6 samples. Multiple NLP models were examined, differing in the type of dictionary components (word length, step, context) as well as the encoding length and number of sequences required for algorithm training. The best configuration analyses the entire sequence and uses a word length of 3 base pairs with one-word neighbor on each side. For the considered FDXR gene, the achieved mean balanced accuracy (BACC) was 98.29% and NPV 99.25%, compared to *minimap2's* performance in a cross-validation scenario. The next stage focused on exploring the dictionary components and attempting to optimize it, employing statistical techniques as well as those relying on the explainability of the decisions made. Reducing the dictionary from 1024 to 145 changed BACC to 96.49% and the NPV to 98.15%. Obtained NLP model, validated on an external independent genome sequencing dataset, gave NPV of 99.64% for complete and 95.87% for reduced dictionary. The salmon-estimated read counts differed from the classical approach on average by 3.48% for the complete dictionary and by 5.82% for the reduced one.

**Conclusions:** We conclude that for long Oxford Nanopore reads, an NLP-based approach can successfully replace classical mapping in case of emergency. The developed NLP model can be easily retrained to identify selected transcripts and/or work with various long-read sequencing techniques. Our results of the study clearly demonstrate the potential of applying techniques known from classical text processing to nucleotide sequences and represent a significant advancement in this field of science.

**Keywords**

# 1. Introduction

Understanding the DNA code and searching for specific sequences within them has been a subject of research for years [15]. It has led to a number of discoveries and innovations, bringing different ways of sequencing the obtained reads [14], to which one then tries to assign an origin. The first ways of reading nucleotides, such as the Sanger method [13] and the Maxam-Gilbert method [12], created a good starting point for development, later extended by the Illumina technology [11]. It is part of the so-called second-generation NGS (Next Generation Sequencing), significantly speeding up the sequencing process. Despite its wide popularity, the limitation of short reads has led to the emergence of TGS (Third Generation Sequencing) methods, one representative of which is the company Oxford Nanopore Technologies (ONT) with its sequencing approach [10]. It is characterized by the ability to read much longer reads, whose average length is measured in thousands of bp (base pairs), compared to hundreds of bp for NGS [9]. The technique proposed by ONT involves passing nucleic acids through nanopores (protein channels), thereby causing changes in the measured electrical signal used for sequence identification. This results in a short sequencing time, while preserving the native form of DNA/RNA [8].

Regardless of the technique used, identifying the occurrence of a given genome fragment is an essential task. This information allows, among other things, the discovery of new treatments and therapies. The current approach is based on the use of a mapping process, which involves comparing a read to a certain reference sequence. Appropriate software, called an aligner, analyses the match between the two nucleotide sequences and determines the most likely location of the read on the reference. The main alignment algorithms include Needleman-Wunsch [24] and Smith-Waterman [25], representing dynamic programming, and one of the available and ready-to-use tools is minimap2 [17, 18], which supports ONT long-reads.

The described way of aligners works has some limitations due to its generality. Performing a full analysis provides the sequence match location, which is not useful for tasks such as classification. In addition, the mapping time itself is related to the length of the sequence, the number of reads, the aligner used and the available computing resources. However, in some situations, the most important information is the occurrence (or not) of the sought-after sequence in a long-read, disregarding the exact location or differences in matching. This paper considers the problem of ionizing radiation, being a permanent element of the environment in today's life, without which the surrounding world is difficult to imagine. Despite the perception as a harmful factor, it occurs in basic medical procedures such as lung X-rays and CT scans. Moreover, radiation is an integral part of nuclear power stations, which have historically experienced various types of accidents (Chernobyl, Fukushima). Unfortunately, it becomes a threat difficult to detect due to invisibility and insensibility. Therefore, research is still ongoing to find potential markers to help in the task. Based on the available literature [6], there are a number of genes that appear to be suitable for biological dosimetry. Our main goal was to propose a 'noMapping mapping' solution to find the sought-after sequence in a set of long-reads and provide an alternative to classical bioinformatics methods. The idea presented was to replace the aligner (and its sequence matching algorithm) with a machine learning model using techniques known from natural language processing (NLP). The classifier created was designed to find ONT long-reads that potentially contain the sought-after sequence. Such filtering allows undesirable reads to be discarded at an early stage and permits further (more detailed) analysis, for example, determining absorbed radiation dose in order to serve as a screening test.

The main NLP technique was the 'bag of words' method, allowing the text to be converted into a vector of numbers that makes possible further calculations. A similar approach was proposed in a paper [5], which focused on finding the viral genome. It used NGS data as the study material and employed de novo genome assembly, distinguishing it from our solution. Another paper [4] used the 'TF-IDF' technique, being an NLP alternative to 'bag of words'. The authors therein focused on detecting regions of lateral origin, relying on the frequency distributions of k-mers in the sequences. In addition, our previous studies [21, 22, 23] have shown the potential to explore the solution described in this paper in more depth.

As far as we are aware, it is the first work analyzing ONT long-reads to identify the sought-after gene, using such NLP encoding and not requiring other time-consuming preprocessing operations. The outcomes shown in this paper focus on the analysis of the FDXR gene, but the approach used can easily be applied to other sought-after sequences as well (confirmed for the NACA gene and described later). A generic implementation of 'noMapping mapping' has been made available at: https://github.com/DrDEXT3R/noMapper.

## 2. Methods

### 2.1. Data

Two datasets, marked as I and II, were used in this study. The first one, RNA sequencing dataset, was utilized for all the numerical experiments performed, both for training and pre-testing the machine learning models (subsection 'Experimental design'). The second was a genome sequencing dataset, with different properties, which was used solely as an independent validation set, thus verifying the final effectiveness of the proposed solution (subsection 'Testing on an external dataset').

### 2.1.1 Dataset I

Dataset I contained Oxford Nanopore long-reads. Full-length sequencing was performed on a GridION sequencer with libraries prepared using the direct RNA sequencing kit (SQK-RNA002). All available data were generated from three repetitions of the biological experiment: A, B, C. They were prepared using the HT1080 cell line and in each repetition the cells in T-25 flasks were exposed to a 10 Gy X-rays dose. Once irradiated, the samples were incubated at 37 °C with 5% $CO_2$ for 24 hours. The cell line was maintained in Minimum Essential Medium (MEM) containing 10% FBS (fetal bovine serum) and 1% penicillin/streptomycin.

In the subsequent stages, RNA extraction was carried out using the RNeasy Mini kit following the manufacturer's instructions. Quantity of isolated RNA was determined by spectrophotometry with a ND-1000 NanoDrop and quality was assessed using a Tapestation 2200. The resulting dataset consisted of six samples, three of which were non-irradiated (A1, B1, C1) and a further three samples 24 hours after exposure (A2, B2, C2). Eventually, ONT sequencing yielded 8.5 million RNA long-reads.

### 2.1.2 Dataset II

Dataset II was an excerpt from ONT's GM24385 open dataset (SRE version after base-calling with Guppy 5.0.6). It contained high molecular weight DNA from lymphoblastoid cells, representing the human genome. However, the present work did not use all the available sequences, but only two chromosomes: one on which the FDXR gene is located (no. 17) and the other randomly selected containing a relatively similar number of sequences (no. 14). More information on the entire shared GM24385 dataset can be found [20].

## 2.2. Data preparation

The sequencing data were subjected to several procedures to prepare them for further analysis. The whole process started with finding and removing possible adapters using Porechop [19]. Filtering was then carried out, thereby removing short reads. This step was performed under the assumption that they could be the result of various errors, which would not have a positive impact on the functioning the entire proposed solution. In order to determine the threshold value, an Empirical Cumulative Distribution Function (ECDF) was drawn for each sample from dataset I. One of these is shown in supplementary Figure S1. A value was chosen as the cut-off threshold for which the beginning of the graph line to its right had a steep slope to the horizontal axis, while not excluding too many reads from further analysis. After detailed comparative analysis across all samples, the cut-off value was set to 500 bp. Therefore, all reads whose length was less than 500bp were discarded. Finally, almost 7.2 million reads remained in the dataset I for further use. For each sample, descriptive statistics of location were calculated, in the form of string length quartiles (Q1, median, Q3). The overall summary is presented in Table 1.

**Table 1.** A number of reads and read length quartiles for a dataset I.

| Sample | Not filtered | Filtered (>500bp) | | | |
|---|---|---|---|---|---|
| | Long-reads | Long-reads | Q1 | Median | Q3 |
| A1 | 1,232,364 | 1,040,293 | 784 | 1231 | 1692 |
| B1 | 1,665,661 | 1,410,171 | 780 | 1204 | 1702 |
| C1 | 1,254,760 | 1,069,828 | 802 | 1263 | 1720 |
| A2 | 1,670,162 | 1,356,675 | 745 | 1146 | 1556 |
| B2 | 1,859,656 | 1,570,158 | 786 | 1229 | 1721 |
| C2 | 862,354 | 718,141 | 772 | 1209 | 1648 |
| **Total** | **8,544,957** | **7,165,266** | | | |

Making a selection of ONT long-reads, still required assigning them to one of two categories: 'gene/transcript' and 'no-gene/transcript'. The first referred to such reads that could potentially belong to a particular genome fragment. The second referred to the opposite case. In order to accomplish the task posed in this way it was decided to use an alignment software, minimap2 [17, 18], which is responsible for aligning the given sequences to a given reference sequence. Using it, finding the likely location of each read becomes possible. Since the need to divide into two groups, the sought-after gene was chosen as the reference sequence. Ultimately, the majority of long-reads (7,128,352; 99.48%) were not mapped. All the rest of reads (36,914), according to the aligner, could potentially come from the gene being searched for.

**2.3. Classifier structure**

The structure of the system intended for classification, consisted of two main components. The first was responsible for appropriately encoding the input sequence and the second for performing the output prediction. The proposed concept employs a well-known and widely used technique from natural language processing, the 'bag of words' [16]. It is based on counting the occurring keywords, belonging to a finite set (dictionary), which allows the input information to be represented as a vector of natural numbers. The algorithm works successfully with 'classic' text containing words, i.e. letter combinations separated by spaces. However, in the DNA/RNA data, there is a problem with defining the 'word' in strings of nucleotides. Splitting the whole sequence into k-mers is a solution. Such words have a fixed length and are formed with a predetermined step, allowing long strings of nucleotides to be interpreted as an NLP problem. The encoded sequences as vectors were subjected to prediction, determining the probability of origin from a chosen gene. A fully connected neural network consisting of 3 layers was used as a classifier. The first two were built of 50 neurons and had ReLU as an activation function. The output layer contained only 1 neuron with a sigmoid function, deciding the membership of one of the two classes.

**2.4 Classifier's degrees of freedom**

The implementation of the presented encoding process requires the setting of certain parameters, which impacts the final classification quality. One example of a value potentially influencing its efficiency and, at the same time, a necessary parameter from the point of view in generating words (k-mers) is their length ($\kappa$). Additionally, having a long sequence of nucleotides, subsequent vocables can be generated with a certain step $\varphi$, meaning an offset from the beginning of the previous one. Apart from that, the finite dictionary needed for encoding may contain components representing single words or also their neighborhood/context ($\lambda$). It is a case of a combination of several consecutive words representing a context, which would, in such a case, be a single dictionary component. Another parameter needed to consider is the impact of the encoded long-read length $\tau$. Perhaps for some cases, analyzing only the initial $\tau$ nucleotides is sufficient, making the calculation certainly faster.

The last aspect is the size and structure of a training dataset ($\Omega$). In this paper, all such parameters are called degrees of freedom, and the significance of which has been analyzed in depth. Figure 1 presents a visualization of the analyzed degrees of freedom.

**Fig. 1.** Considered degrees of freedom.

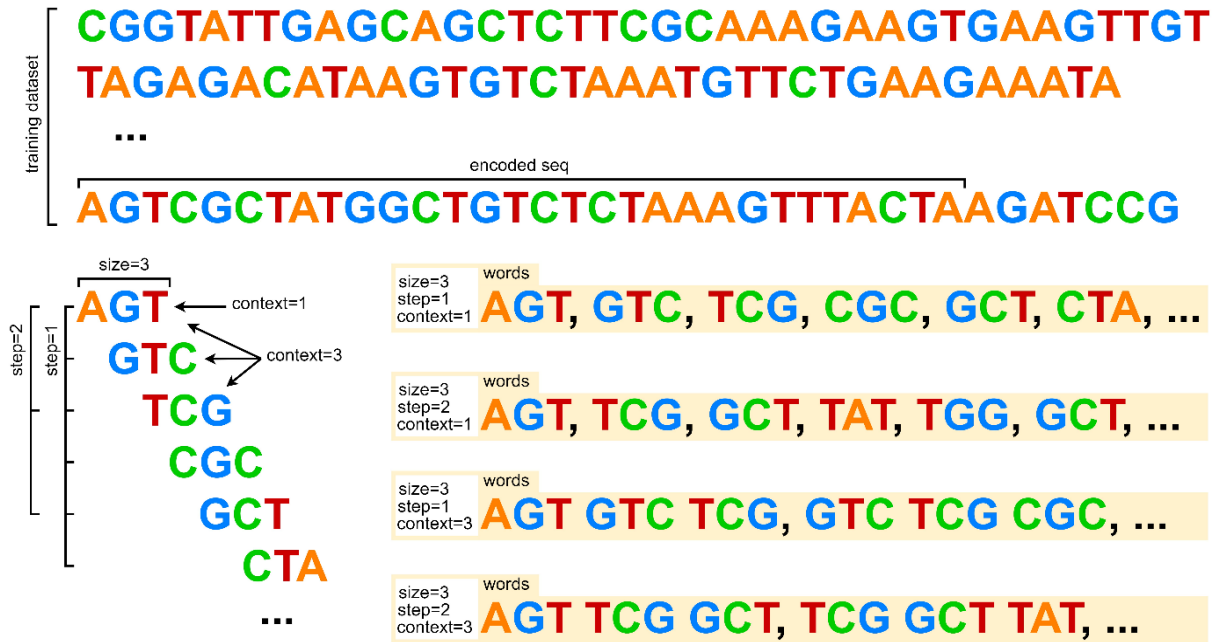

## 2.5 Experimental design

The main aim of the study is to construct an NLP-based classification system that allows the detection of sequences that could potentially originate from the gene being searched for. Several configurations of parameters were analyzed in the leave-one-sample-out-cross-validation (LOSOCV) schema. Firstly, the randomly chosen sequences from both categories (gene and no-gene transcripts, with a 1:3 ratio to emphasize the rarity of the first group) from each sample constituted six fixed testing datasets, one for each LOSOCV experiment.

A preliminary configuration setting experiment focused on the subset of parameters. Reflecting the genetic code, the word length was initially assumed to mimic one codon and to be equal to 3bp ($\kappa = 3$). Next, to see which approach would perform better for the classification task, a comparison was made between a dictionary containing single words ($\lambda = 1$) and a dictionary built from words with context, built with one 3bp neighbor on each side ($\lambda = 3$; 3bp vs. 3bp|3bp|3bp). For this experiment, step $\varphi = 1$ was used and all nucleotides of each sequence were encoded ($\tau$ = whole seq). The first parameter investigated for its impact on the final prediction performance was the training set size $\Omega$. Since the task of finding a sequence that would potentially be derived from a specific gene, it was decided to use an unbalanced dataset at the training stage. The criterion was based on biological reality, as the sequence sought represents only a small fraction of the entire genome. The assumed ratio of representatives of the two classes was 1:3, where the second referred to sequences other than the gene searched for (just like the test set). Based on this assumption, the following datasets were considered: $\Omega$ = {1,000 + 3,000; 3,000 + 9,000; 10,000 + 30,000; ~30,000 + 90,000; ~30,000 + 300,000 and ~30,000 + 900,000}. To clarify, ~30,000 should be understood as all available sequences derived from the gene, the exact number of which varies according to the cross-validation experiment. Therefore, the imbalance ratio in subsequent training sets increases. Additionally, in order to shorten the notation, '~' was omitted and the numeral k was used to denote a thousand.

Having the results from the preliminary experiment, it was decided to select such a training dataset $\Omega$ and neighborhood $\lambda$ for which the classifier achieved the best performance, and then to carry out the main experiment, allowing the remaining degrees of freedom to be analyzed. The number of nucleotides undergoing encoding $\tau$ was taken as the whole sequence and the values representing measures of position - string length quartiles: Q1, median, Q3. Then, observing a constant difference between these values, it was decided to augment this set with a number even lower than Q1. In the end, five different values $\tau$ were obtained. Deciding on a two-element set of steps $\varphi \in \{1,2\}$ was supported by the occurrence of mutations and, in general, the appearance of read errors, which is particularly characteristic of ONT long-reads. However, it is worth mentioning that as the parameter $\varphi$ increases for $\lambda > 1$, the size of the final dictionary for the bag-of-words method rises. This is due to the higher number of permutations, which translates into the memory complexity of the hardware used to train the model. Therefore, among other reasons, it was decided to consider maximally two values $\varphi$. As the last degree of freedom determining the length of the k-mer, $\kappa \in \{3,6\}$ was chosen. Again, the first value is related to the amino acid encoding and the second value is its doubling. Similar to the $\varphi$ parameter, the size of $\kappa$ directly impacts the cardinality of the final dictionary. By selecting one value $\lambda$ and $\Omega$, the final sets $\tau$, $\varphi$, $\kappa$, it was possible to conduct the main experiment and analyze the effectiveness of all classification models. Finally, 20 classifiers were compared with each other.

## 2.6 Testing on an external dataset

Using the conclusions of the experiments, it became possible to select the optimal configuration of parameters. Attempts were made to choose values that would allow the creation of an efficient classifier with the smallest possible dictionary size. Then, the trained model on dataset I, was verified on dataset II. It is worth noting that the second dataset is completely dissimilar, is genome not transcriptome sequencing, comes from a separate source, has different features, therefore, the tests carried out answered how effective the proposed solution is. The generality property of classifiers is particularly important because it determines the effectiveness of models on previously unseen input data.

## 2.7 Dictionary optimization

The size of the dictionary depends on the configuration parameters, and its range can include significant values depending on the choices made. Consequently, a key aspect becomes the optimization problem, which involves selecting the most relevant words in order to effectively manage the dictionary size. In the context of the case presented, certain nucleotide sequences (containing certain words) occur significantly more frequently in the analyzed gene compared to the rest of the genome. Moreover, they can be unique and only occur in one location, which is crucial in the classification task.

Taking these considerations into account, it was decided to examine the importance level (weight) of each dictionary component. The features were ranked in ascending order, from the least important to the most important ones – a component with the least weight was assigned the number one, while the most relevant dictionary component was placed at the end of the list with the highest possible rank. The weights (feature importance factors) were assigned based on three different approaches: odds ratio (OR) based, effect size based and using explainable artificial intelligence (XAI) tools. For the first, OR-based method, a 1/OR transformation was applied for OR values less than 1, which makes feature rank independent from its type of impact (risk or protection). The effect size-based method used Cramer's V [1]. In the third method, the SHAP tool [3] was used. Finally, three independent rankings of features were obtained, which were later collated together and their selection consistency compared.

The optimization task was completed by investigating the impact of reducing the dictionary size on the efficiency level of the models. In a first step, dominant words were selected, meaning those with the highest weights, and a reduced feature space was used to train the model. The selection of dominant words can be approached in several ways. The simplest one is based on choosing a fixed *d*, which determines the number of words with the highest weight. This approach is quite complex, namely the question arises what value of *d* should be chosen. To solve the problem, two different data-driven approaches, known from the analysis of Receiver Operating Characteristic (ROC) in machine learning optimization task, were used. The previously calculated word importance values were sorted and plotted to constitute the importance curve. In the first method, called 'max distance', the distance of each importance curve data point from a straight line connecting the first and last points was calculated. The data point (representing a particular component of the dictionary) with the maximum distance defined the cut-off. All words with an importance score higher than that were included in the reduced dictionary. The second approach uses piecewise linear regression to model the importance curve. The dividing points were determined by minimizing the residual sum of squares for the entire dataset. The cut-off point was the data point separating the first and second regression lines (see supplementary Figure S2).

## 2.8 Evaluation metrics

The approach developed requires tuning several parameters, which was done stepwise. Firstly, we selected the values of $\lambda$ and $\Omega$ to define the and then the remaining configuration parameters that determine the most effective classifier. Several indicators can evaluate a classifier, each focusing on a different property to determine the model's effectiveness. As the focus is on the sequence filtering task, the main objective is to separate all potentially possible gene sequences from the rest. Omitting a read that potentially contains the specific gene is expected to occur rarely. Therefore, the negative predictive value (NPV), defined as the fraction of true negatives among all negatives obtained, is chosen to be maximized. To determine the confidence that the reads classified into the group of the searched sequence actually contained it, we used PPV (positive predictive value), which represents the fraction of true positives among all positives. Additionally, we used BACC (balanced accuracy) to evaluate the overall effectiveness of the proposed solution, due to the existing imbalance among the classes.

## 3. Results and discussion

As mentioned earlier, the whole study included two experiments: a preliminary and a main experiment. The first one allowed to select the training set size $\Omega$ and the neighborhood value $\lambda$, while the second one allowed to compare different NLP encoding configurations: $\tau, \varphi, \kappa$.

## 3.1 Preliminary experiment

Figure 2 shows the balanced accuracy of the classifiers for which the training set size was altered.

The approach with single words ($\lambda = 1$) in the dictionary and one with two-sided neighbors ($\lambda = 3$) were compared to each other. In turn, the exact NPV and PPV values, together with the number of representatives in the final dictionary for the same sets, are summarized in Table 2. Assumed values for the remaining parameters are: $\tau = whole\ seq$, $\varphi = 1$, $\kappa = 3$. The results clearly illustrate the effectiveness of the applied method for the classification task. Even for a relatively small dataset, the model is able to classify sequences with an average of 87.22% ($\lambda = 1$) and 95.11% ($\lambda = 3$) balanced accuracy. The most effective model expressed by this indicator was achieved for $\Omega = 30k + 90k$ (all reads potentially derived from the sought-after gene), which had a very high value of 98.29% for words with the neighborhood, PPV = 96.58% and NPV = 99.25%. The one on the same set $\Omega$, also proved to train the classifier most effectively considering models with $\lambda = 1$, where BACC = 92.35%,

PPV = 92.47%, and NPV = 95.78%. By directly comparing the models against the considered parameter λ, it can be concluded that the use of neighborhood words improves the effectiveness of the classifier by about 5-9% for balanced accuracy and 3-5% for NPV. However, the nature of the changes depending on Ω is similar. Initially, there is a dynamic improvement in efficiency up to a saturation level, and then worse model evaluation rates are achieved. However, the advantage of the classifiers based on λ = 1 was the very small size of the final dictionary. Due to the four-nucleotide alphabet, the number of possible permutations of three-letter words was only 64. This was as much as sixteen times less than classifiers with a neighborhood of λ = 3.

**Fig. 2.** Balanced accuracy (BACC) depending on training set size for 3bp and 3bp|3bp|3bp.

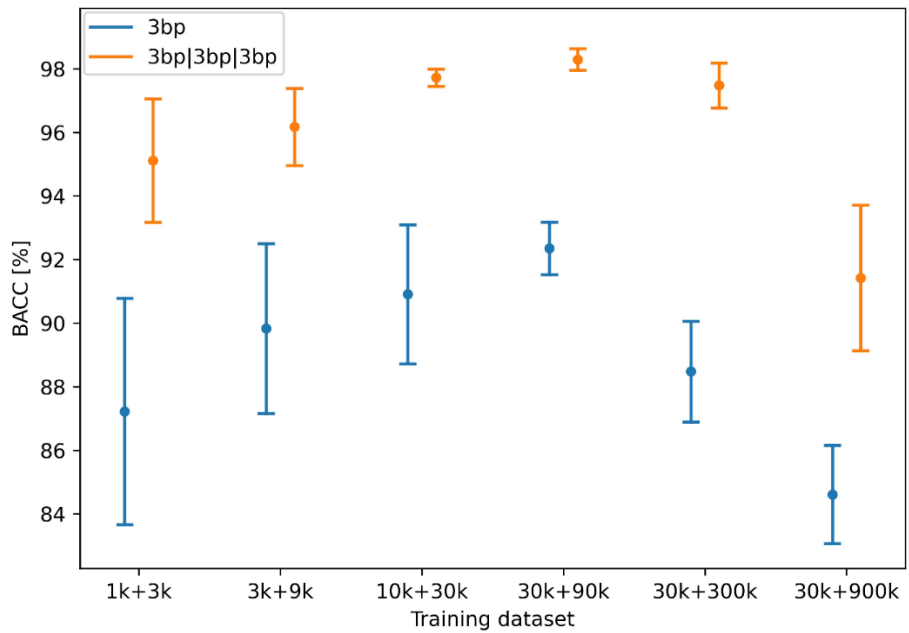

*Results are presented as mean values and its 95% confidence intervals. 3bp is λ = 1 and 3bp|3bp|3bp is λ = 3.*

**Table 2.** Mean values of NPV, PPV and BACC with their 95% confidence intervals (in brackets).

| Dataset size | Indicator | 3bp [%] | 3bp\|3bp\|3bp [%] |
|---|---|---|---|
| 1k+3k | NPV | 93.84 (90.11, 97.56) | 97.44 (95.89, 98.99) |
|  | PPV | 84.80 (71.22, 98.38) | 94.31 (90.53, 98.10) |
|  | BACC | 87.22 (83.66, 90.78) | 95.11 (93.17, 97.05) |
| 3k+9k | NPV | 94.32 (92.51, 96.13) | 97.98 (96.92, 99.05) |
|  | PPV | 91.46 (88.67, 94.25) | 95.54 (92.24, 98.85) |
|  | BACC | 89.83 (87.16, 92.50) | 96.17 (94.96, 97.38) |
| 10k+30k | NPV | 95.02 (93.29, 96.76) | 98.95 (98.61, 99.30) |
|  | PPV | 91.63 (86.12, 97.14) | 95.82 (94.05, 97.59) |
|  | BACC | 90.91 (88.72, 93.09) | 97.72 (97.45, 97.99) |
| 30k+90k | NPV | 95.78 (95.10, 96.47) | 99.25 (98.94, 99.55) |
|  | PPV | 92.47 (90.29, 94.66) | 96.58 (95.72, 97.44) |
|  | BACC | 92.35 (91.53, 93.18) | 98.29 (97.95, 98.63) |
| 30k+300k | NPV | 92.99 (92.07, 93.91) | 98.62 (98.12, 99.11) |
| | PPV | 97.96 (97.56, 98.36) | 97.39 (96.32, 98.47) |
| | BACC | 88.48 (86.89, 90.06) | 97.48 (96.77, 98.18) |
| 30k+900k | NPV | 90.73 (89.88, 91.58) | 94.65 (93.26, 96.04) |
| | PPV | 99.39 (98.83, 99.96) | 99.41 (99.12, 99.69) |
| | BACC | 84.61 (83.06, 86.16) | 91.42 (89.13, 93.71) |
| | | *Dictionary size = 64* | *Dictionary size = 1,024* |

## 3.2 Main experiment

After a preliminary experiment, the parameters $\Omega = 30k + 90k$, $\lambda = 3$ were chosen, and the next experiment focused on the impact of word length and step. The number of the first base pairs from the analyzed sequence was also parametrized. The obtained estimates of balanced accuracy are shown in Figure 3.

**Fig. 3.** Model balanced accuracy (BACC) depending on the encoded sequence length.

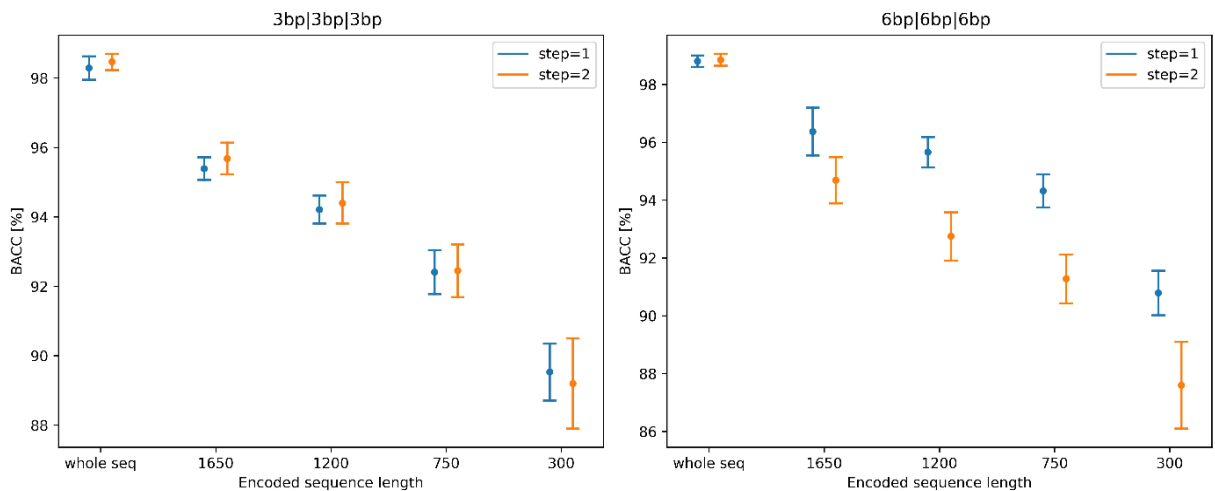

*Steps $\varphi \in \{1; 2\}$ and $\kappa = 3$ (left panel) and $\kappa = 6$ (right panel) were used. Results are presented as mean values and their 95% confidence interval.*

They compare models varying in the number of encoded nucleotides $\tau$ and the word generation step $\varphi$. In the left panel, the results for $\kappa = 3$ (3bp|3bp|3bp) are accumulated, and in the right one for $\kappa = 6$ (6bp|6bp|6bp). The NPVs, PPVs and BACCs are summarised in supplementary Table S1 and Table S2. Again, mean values and their 95% confidence interval based on all LOSOCV replicates are used.

As expected, the best performance was achieved by classifiers that encoded all nucleotides of long-reads. Regardless of $\varphi$ and $\kappa$, their BACC was over 98%, PPV over 96% and NPV over 99%. Analyzing Figure 3 (left panel) and Table S1, it can be deduced that for the 3bp|3bp|3bp encoding there was no significant improvement in

model efficiency depending on the step $\varphi$. However, there was a significant increase in the size of the final dictionary. A sixteen-fold larger list, translated into significantly higher memory complexity, but did not result in noticeable increases in classification. The situation was different for the 6bp|6bp|6bp encoding (Figure 3, right panel), where, comparing the models against the parameter $\varphi$, those with a lower step were characterized by a better balanced accuracy score for all tested values of $\tau$ other than whole seq. The same situation is observed for the NPV indicator, shown in Table S2. It is also worth noting the significant size of the final dictionary for $\varphi = 2$. The number of permutations was so large that only for such parameters configuration, not all possible combinations of 6bp|6bp|6bp were included in the dictionary. Nevertheless, encoding all the nucleotides of long-reads in this way achieved the best classifier considering BACC and NPV indicators. The worst among the $\tau = $ whole seq models was the one with 3bp|3bp|3bp encoding for $\varphi = 1$. However, the difference in the quality measures analyzed is marginal, especially bearing in mind the approximately 1017% smaller size of the final dictionary compared to the most effective of them.

**3.3 Testing on an external dataset**

As mentioned before, performing the preliminary and main experiments made it possible to select the optimal values for the configuration parameters. Based on the results presented, it was decided that the classifier tested on the external dataset should have the following properties: neighborhood $\lambda = 3$, word length $\kappa = 3$ and step $\varphi = 1$. $\Omega = 30k + 90k$ was chosen as the training set and $\tau = $ *whole seq* was encoded. This configuration provided a 1024 components dictionary, with high model efficiency. Running a verification test on an external dataset II provided feedback for which the BACC = 75.28%, PPV = 66.96% and NPV = 99.64%. These results were based on 58,574 long sequences, with 29,287 representatives in each class.

It can be seen that, despite a completely different dataset, characterized by dissimilar properties from dataset I, the results confirmed the effectiveness of the proposed solution. Noteworthy is the very high value of the NPV metric, which is especially important from the point of view of the considered task and further analysis of sequences potentially derived from the searched gene. Thus, it seems that the 3bp|3bp|3bp encoding, with dictionary size = 1024, fulfils the task of filtering out long-reads that with high confidence do not originate from the sought-after gene.

## 3.4 Dictionary optimization

Using exactly the same configuration parameters, a dictionary containing 1024 components was analyzed. It began by creating three independent rankings, which were then compared. The rank value represented the feature importance – the higher ranking value the more important component ('word') is.

First, each component was assigned a position in all rankings, and next the average position was calculated, which was used to order the dictionaries. In this way, supplementary Figure S3 and Figure S4 were created, focusing on the components with the highest ranks. The green line indicates the perfect match of all three methods.

**Table 3.** Dictionary size after reducing its components to dominant ones for each LOSOCV repetition.

| Selection strategy | Max distance | | | | | | Regression-based | | | | | |
|---|---|---|---|---|---|---|---|---|---|---|---|---|
| | A1 | B1 | C1 | A2 | B2 | C2 | A1 | B1 | C1 | A2 | B2 | C2 |
| Odds ratio | 149 | 156 | 145 | 170 | 173 | 169 | 143 | 139 | 143 | 140 | 135 | 142 |
| Effect size | 117 | 118 | 118 | 110 | 114 | 120 | 88 | 93 | 88 | 93 | 95 | 87 |
| XAI | 74 | 53 | 77 | 73 | 52 | 91 | 44 | 44 | 52 | 55 | 44 | 41 |

The next step was to select the most dominant features treated as gene (transcript) fingerprints. Using the max distance and regression approaches, cut-off points were determined, i.e. the number of dictionary components with the highest significance level. The sizes of the reduced dictionaries are presented in Table 3. Subsequently, the consistency in selection of such dictionaries was compared. First against all ranking methods per each LOSOCV (supplementary Figure S5 for exemplary LOSOCV) and then across all LOSOCV repetitions (Table 4).

**Table 4.** Consistency in selecting dominant dictionary components across LOSOCV repetitions in relation to the first repetition.

| | Max distance | | | Regression-based | | |
|---|---|---|---|---|---|---|
| LOSOCV | OR | Effect size | XAI | OR | Effect size | XAI |
| A1 - reference | 1.0000 (149) | 1.0000 (117) | 1.0000 (74) | 1.0000 (143) | 1.0000 (88) | 1.0000 (44) |
| B1 | 0.9574 (146) | 0.9702 (114) | 0.6457 (41) | 0.9574 (135) | 0.9724 (88) | 0.6591 (29) |
| C1 | 0.9864 (145) | 0.9872 (116) | 0.6623 (50) | 0.9860 (141) | 0.9886 (87) | 0.7292 (35) |
| A2 | 0.9342 | 0.9604 | 0.6667 | 0.9894 | 0.9613 | 0.6263 |
|    | (149)  | (109)  | (49)   | (140)  | (87)   | (31)   |
| B2 | 0.9255 | 0.9524 | 0.6349 | 0.9712 | 0.9508 | 0.6136 |
|    | (149)  | (110)  | (40)   | (135)  | (87)   | (27)   |
| C2 | 0.9371 | 0.9873 | 0.6424 | 0.9895 | 0.9943 | 0.7059 |
|    | (149)  | (117)  | (53)   | (141)  | (87)   | (30)   |

*The Dice similarity coefficient is shown, along with the number of common dictionary components (in brackets).*

The analysis was concluded by comparing how the dictionary restriction affects the final performance of the classifiers. During this step, the focus was on comparing the 95% confidence intervals for NPV, PPV, BACC. The obtained results are presented in Figure 4.

**Fig. 4.** The summarized classifier's performance in LOSOCV obtained for different dictionary optimization strategies.

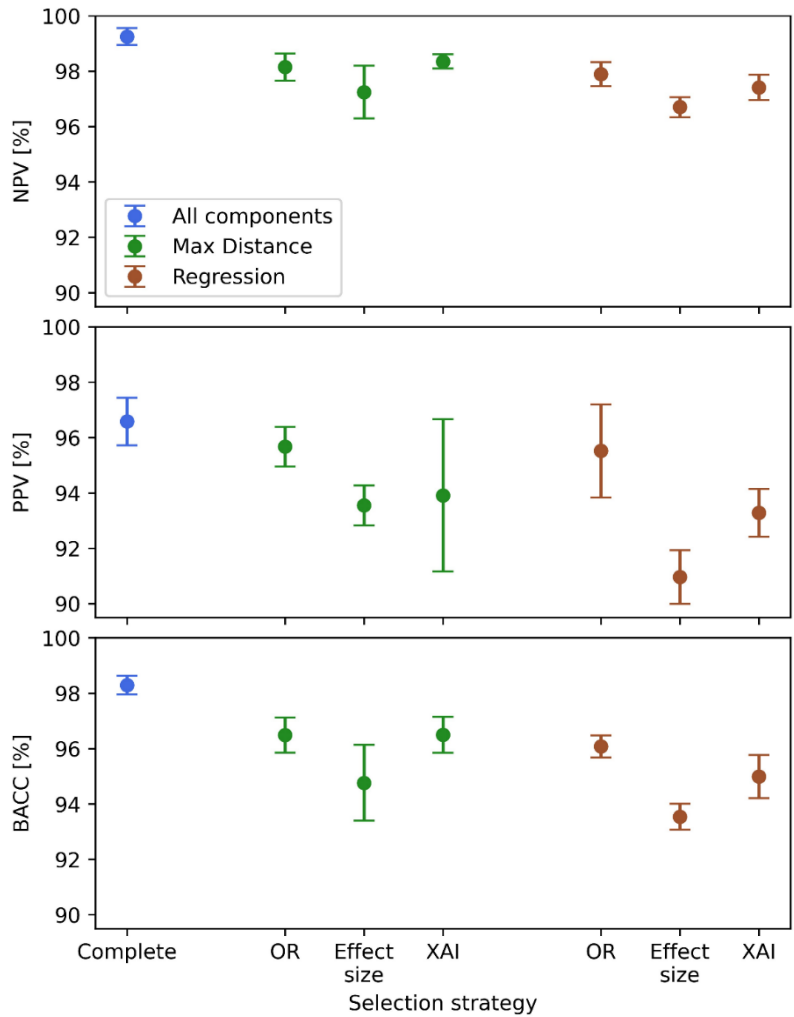

*Results are presented as mean value and its 95% confidence interval.*

Based on the results obtained, there is a consistency in the selection of the most significant dictionary components. Regardless of the chosen ranking method, such words needed for encoding are outstanding and turn out to be of clear relevance during model prediction. This is shown in supplementary Figure S3 and Figure S4. At first, for the initial dictionary components, there is quite a spread of differences in ranking positions, but as one approaches the words with the highest importance, the spread decreases and the points tend towards the line of perfect match.

In the case of methods for selecting dominant features, it can be seen (Table 3) that the regression approach generally selects fewer dictionary components, and the difference from max distance depends on the nature of the plot (related to the way the weights were calculated). The least noticeable difference is for the OR-based method. However, it is important to say that the components in the reduced dictionaries are repeated. The phenomenon occurs across the LOSOCV repetition (supplementary Figure S5) and in the ranking methods (Table 4). The XAI-based approach achieves the lowest values of the Dice similarity coefficient, but it is worth noting that the least numerous dictionaries of dominant features are also observed. The other two approaches observe high coverage of selected components.

Importantly, the NPV metric decreases relatively slightly (Figure 4), despite reducing the dictionary components by up to 20 times (XAI & regression). For the OR-based method, the average NPV dropped from 99.25% (all components) to 98.15% in the max-distance-based optimized dictionary and 97.89% in the regression-based. For effect size, averages of 97.24% (max distance) and 96.70% (regression) were achieved. Similarly, NPV fell to 98.35% and 97.41% for the XAI-based method. Such a property allows increased efficiency during the encoding stage and reduced model prediction time. The worst performance was achieved by classifiers with dominant features selected using effect size. The OR-based technique performed best, but it should be noted the classifiers were trained on the largest feature space (reduced dictionary size: 135-173). PPV values generally reached lower levels compared to NPV, but a similar nature of change can be observed. The highest PPV = 96.58% reached the complete dictionary. The max distance method resulted in decreases to 95.67% (OR), 93.55% (effect size) and 93.91% (XAI), while the regression technique led to 95.52% (OR), 90.96% (effect size) and 93.28% (XAI). It is noteworthy that in all cases, except the effect size & max distance combination, a wider confidence interval is observed in comparison to NPV. In the last indicator also the greatest BACC = 98.29% was observed for the complete dictionary. Methods based on OR, effect size and XAI achieved: 96.49%, 94.76%, 96.50% for max distance and 96.08%, 93.54%, 94.99% for regression.

The final evaluation of the developed no-mapping sequence aligner used the gene expression values as indicator. All sequences classified as gene/transcript related were subjected to transcript counts per million estimation by *salmon* software [2]. NCBI database was chosen for gene transcript references. The results are presented in Table 5. Additionally, to verify the effectiveness of the proposed solution, the same approach was performed for the NACA gene, characterized by higher expression. The results are included in Table S3.

**Table 5.** Standardized FDXR transcript/gene count estimates (per million) for different data processing models.

| Transcript | A1 | A2 | B1 | B2 | C1 | C2 |
|---|---|---|---|---|---|---|
| Classical approach: minimap2 + salmon | | | | | | |
| NR_047576.3 | 0 | 24.31 | 15.29 | 0 | 0 | 0 |
| NM_001258014.4 | 13.79 | 0 | 0 | 0 | 0 | 0 |
| NM_024417.5 | 13.79 | 156.46 | 15.47 | 110.16 | 28.29 | 59.03 |
| NM_004110.6 | 13.79 | 0 | 15.47 | 91.08 | 28.29 | 92.59 |
| NM_001258015.3 | 0 | 0 | 0 | 0 | 0 | 0 |
| NM_001258012.4 | 0 | 4.24 | 0 | 0 | 0 | 21.50 |
| NM_001258013.4 | 0 | 0 | 0 | 24.07 | 0 | 33.28 |
| NM_001258016.3 | 0 | 0 | 0 | 0 | 0 | 0 |
| **FDXR total** | **41.38** | **185.01** | **46.23** | **225.31** | **56.58** | **206.41** |
| noMapping mapping (complete dictionary) + salmon | | | | | | |
| NR_047576.3 | 0 | 25.87 | 15.10 | 0 | 0 | 0 |
| NM_001258014.4 | 13.25 | 0 | 0 | 0 | 0 | 0 |
| NM_024417.5 | 13.25 | 149.63 | 15.26 | 103.66 | 27.50 | 56.52 |
| NM_004110.6 | 13.25 | 0 | 15.26 | 84.88 | 27.50 | 89.19 |
| NM_001258015.3 | 0 | 0 | 0 | 0 | 0 | 0 |
| NM_001258012.4 | 0 | 4.12 | 0 | 0 | 0 | 20.60 |
| NM_001258013.4 | 0 | 0 | 0 | 23.33 | 0 | 31.98 |
| NM_001258016.3 | 0 | 0 | 0 | 0 | 0 | 0 |
| **FDXR total** | **39.76** | **179.62** | **45.63** | **211.87** | **54.99** | **198.29** |
| noMapping mapping (reduced dictionary) + salmon | | | | | | |
| NR_047576.3 | 0 | 24.75 | 14.71 | 0 | 0 | 0 |
| NM_001258014.4 | 13.25 | 0 | 0 | 0 | 0 | 0 |
| NM_024417.5 | 13.25 | 143.14 | 14.86 | 101.81 | 25.50 | 57.60 |
| NM_004110.6 | 13.25 | 0 | 14.86 | 82.90 | 25.50 | 90.65 |
| NM_001258015.3 | 0 | 0 | 0 | 0 | 0 | 0 |
| NM_001258012.4 | 0 | 3.95 | 0 | 0 | 0 | 20.98558 |
| NM_001258013.4 | 0 | 0 | 0 | 22.85 | 0 | 32.54 |
| NM_001258016.3 | 0 | 0 | 0 | 0 | 0 | 0 |
| **FDXR total** | **39.76** | **171.84** | **44.43** | **207.57** | **51.01** | **201.77** |

One can notice that the estimation error, as related to the classical approach, varies from 1.30% to 5.97% (mean 3.48%, standard deviation 1.56%) for FDXR gene expression for complete dictionary NLP model and from 2.25% to 9.86% (mean 5.82%, standard deviation 2.91%) for the model using the reduced dictionary. In the case of the NACA gene, the estimation error is 0.55-2.02% (complete dictionary, mean 1.28%, standard deviation 0.59%). Keeping in mind that the targeted approach is the radiation accident victim triage, the obtained accuracy fulfills the system requirements.

## 4. Conclusions

This paper presents a method for classifying long-reads in search of a specific transcript sequence(s). The proposed solution consisted of two main components: the first was responsible for encoding the sequence using NLP methods, and the second was a neural network for performing the sequence classification. Various parameters were analyzed to encode the set of long-reads, as well as construct the training dataset. A total of 31 machine-learning models were considered. The best-performing classifier used the training dataset with 1:3 ratio between possible gene and no-gene categories. The prepared comparison unequivocally showed the advantage of encoding taking into account neighbors over a dictionary containing only single words with a length ($\kappa$) of 3bp. Based on the results obtained, there is a marginal to small effect of the step $\varphi$ parameter on NPV, PPV and BACC. As the parameters $\kappa$ and $\varphi$ rise, the size of the final dictionary increases significantly. During the entire work carried out, the classifier systems ranged from 64 to approximately 1,041,599 encoding elements. This translates into memory and time complexity. Speeding up the encoding process can be done by choosing an suitable value for the parameter $\tau$. As the obtained results showed, when the first 1650 initial nucleotides (equivalent to Q3) were encoded, the decrease in quality of the BACC metric was 2-4% and the NPV 2-3% relative to $\tau$ = whole seq. For 750 initial nucleotides (Q1 equivalent), the quality drops were 4-7% and 3-5%, respectively.

The finally selected classifier system with the configuration of the parameters: $\Omega$ = 30k + 90k, $\lambda$ = 3, $\kappa$ = 3, $\varphi$ = 1 and $\tau$ = whole seq, was tested on an external genome sequencing dataset and the obtained results confirmed the effectiveness of the proposed solution.

Further investigations of classifiers with a dictionary equal to 1024 showed the potential for optimization. The results obtained clearly indicate that some features have a greater or lesser influence on the final prediction of the model. Regardless of the method used to calculate the weights, it is possible to distinguish the components which rank the most influential positions in the rankings. By reducing the dictionaries to only the key ranking places, the effectiveness of the classifiers decreases by 1-3% for NPV, gaining 6-25x smaller dictionary size, depending on the approach used to calculate the weights and locate the cut-off point.

The analysis results presented in this paper show the potential of applying techniques known from NLP to the field of bioinformatics. Appropriate processing of long strings of nucleotides, allows the reads to be treated as 'classic' text, consisting of single words. The demonstrated solution thus provides an alternative to the classical alignment tool. By narrowing down the task to the search for a specific sequence, we can bypass the mapping process and at the same time apply the shown machine learning based method. The proposed 'noMapping mapping' approach can be easily used to identify sequences of interest. High efficiency results have proven the point of transforming

DNA/RNA data into a form friendly to NLP techniques and make a significant contribution to the development of this branch of science.

**Supplementary materials**

**Table S1.** Mean NPV, PPV, and BACC with their 95% confidence interval for encoding 3bp|3bp|3bp.

| Encoded seq length | Step = 1 [%] | Step = 2 [%] |
|---|---|---|
| whole seq | NPV = 99.25 (98.94, 99.55)<br>PPV = 96.58 (95.72, 97.44)<br>BACC = 98.29 (97.95, 98.63) | NPV = 99.33 (99.1, 99.55)<br>PPV = 96.89 (96.24, 97.54)<br>BACC = 98.47 (98.23, 98.70) |
| 1650 | NPV = 97.36 (97.17, 97.55)<br>PPV = 96.28 (95.23, 97.33)<br>BACC = 95.39 (95.07, 95.72) | NPV = 97.55 (97.30, 97.80)<br>PPV = 96.31 (95.75, 96.88)<br>BACC = 95.68 (95.23, 96.14) |
| 1200 | NPV = 96.66 (96.41, 96.91)<br>PPV = 95.67 (94.96, 96.39)<br>BACC = 94.21 (93.81, 94.62) | NPV = 96.74 (96.39, 97.09)<br>PPV = 96.17 (95.09, 97.25)<br>BACC = 94.40 (93.81, 95.00) |
| 750 | NPV = 95.81 (95.32, 96.29)<br>PPV = 92.55 (91.18, 93.91)<br>BACC = 92.41 (91.78, 93.04) | NPV = 95.50 (95.05, 95.95)<br>PPV = 96.28 (95.56, 97.00)<br>BACC = 92.45 (91.69, 93.21) |
| 300 | NPV = 94.49 (93.85, 95.13)<br>PPV = 87.07 (85.07, 89.07)<br>BACC = 89.53 (88.71, 90.35) | NPV = 93.75 (92.92, 94.58)<br>PPV = 93.21 (91.57, 94.85)<br>BACC = 89.20 (87.90, 90.50) |
| | *Dictionary size = 1,024* | *Dictionary size = 16,384* |

*Different step $\varphi \in \{1; 2\}$ and $\Omega = 30k+90k$ were used.*

**Table S2.** Mean NPV, PPV, and BACC with their 95% confidence interval for encoding 6bp|6bp|6bp.

| Encoded seq length | Step = 1 [%] | Step = 2 [%] |
|---|---|---|
| whole seq | NPV = 99.58 (99.40, 99.76)<br>PPV = 96.68 (95.94, 97.42)<br>BACC = 98.81 (98.61, 99.00) | NPV = 99.60 (99.44, 99.76)<br>PPV = 96.76 (96.37, 97.16)<br>BACC = 98.85 (98.65, 99.06) |
| 1650 | NPV = 97.92 (97.36, 98.48)<br>PPV = 97.12 (96.58, 97.66)<br>BACC = 96.37 (95.55, 97.20) | NPV = 96.73 (96.21, 97.25)<br>PPV = 98.29 (97.88, 98.69)<br>BACC = 94.69 (93.89, 95.49) |
| 1200 | NPV = 97.49 (97.16, 97.82)<br>PPV = 96.75 (96.41, 97.10)<br>BACC = 95.66 (95.13, 96.18) | NPV = 95.51 (94.98, 96.03)<br>PPV = 98.49 (98.06, 98.91)<br>BACC = 92.75 (91.91, 93.58) |
| 750 | NPV = 96.67 (96.29, 97.04)<br>PPV = 96.36 (95.61, 97.12)<br>BACC = 94.32 (93.75, 94.89) | NPV = 94.62 (94.11, 95.14)<br>PPV = 98.34 (97.88, 98.79)<br>BACC = 91.28 (90.43, 92.12) |
| 300 | NPV = 94.52 (94.05, 95.00)<br>PPV = 95.71 (94.68, 96.74)<br>BACC = 90.79 (90.02, 91.56) | NPV = 92.52 (91.63, 93.41)<br>PPV = 97.40 (96.78, 98.02)<br>BACC = 87.60 (86.10, 89.10) |
| | *Dictionary size = 65,536* | *Dictionary size = ~1,041,599* |

*Different step $\varphi \in \{1; 2\}$ and $\Omega = 30k+90k$ were used.*

**Table S3.** Standardized NACA transcript/gene count estimates (per million) for different data processing models.

| Transcript | A1 | A2 | B1 | B2 | C1 | C2 |
|---|---|---|---|---|---|---|
| Classical approach: minimap2 + salmon | | | | | | |
| NM_001113202.2 | 60.71 | 23.54 | 18.70 | 51.97 | 15.16 | 0 |
| NM_001320194.2 | 6.93 | 5.83 | 0 | 4.82 | 1.46 | 2.11 |
| NM_001113203.3 | 0 | 0 | 0 | 0 | 0 | 0 |
| NM_001365896.1 | 1.21 | 0 | 0 | 0 | 0 | 0 |
| NM_001320193.2 | 72.53 | 23.54 | 18.70 | 51.97 | 15.16 | 0 |
| NM_005594.6 | 436.29 | 427.98 | 592.47 | 399.50 | 490.51 | 416.32 |
| NM_001113201.3 | 564.85 | 526.21 | 592.47 | 498.91 | 640.48 | 532.46 |
| **NACA total** | **1142.52** | **1007.09** | **1222.34** | **1007.18** | **1162.77** | **950.89** |
| noMapping mapping (complete dictionary) + salmon | | | | | | |
| NM_001113202.2 | 59.39 | 23.18 | 18.47 | 52.27 | 14.98 | 0 |
| NM_001320194.2 | 6.82 | 4.05 | 0 | 3.54 | 1.46 | 2.10 |
| NM_001113203.3 | 0 | 0 | 0 | 0 | 0 | 0 |
| NM_001365896.1 | 1.20 | 0 | 0 | 0 | 0 | 0 |
| NM_001320193.2 | 71.22 | 23.18 | 18.47 | 52.27 | 14.98 | 0 |
| NM_005594.6 | 428.70 | 418.95 | 588.79 | 423.07 | 487.47 | 408.21 |
| NM_001113201.3 | 557.33 | 517.37 | 588.79 | 464.20 | 637.50 | 524.33 |
| **NACA total** | **1124.67** | **986.73** | **1214.53** | **995.34** | **1156.40** | **934.65** |

**Fig. S1.** Empirical Cumulative Distribution Function plot for A1 sample from dataset I.

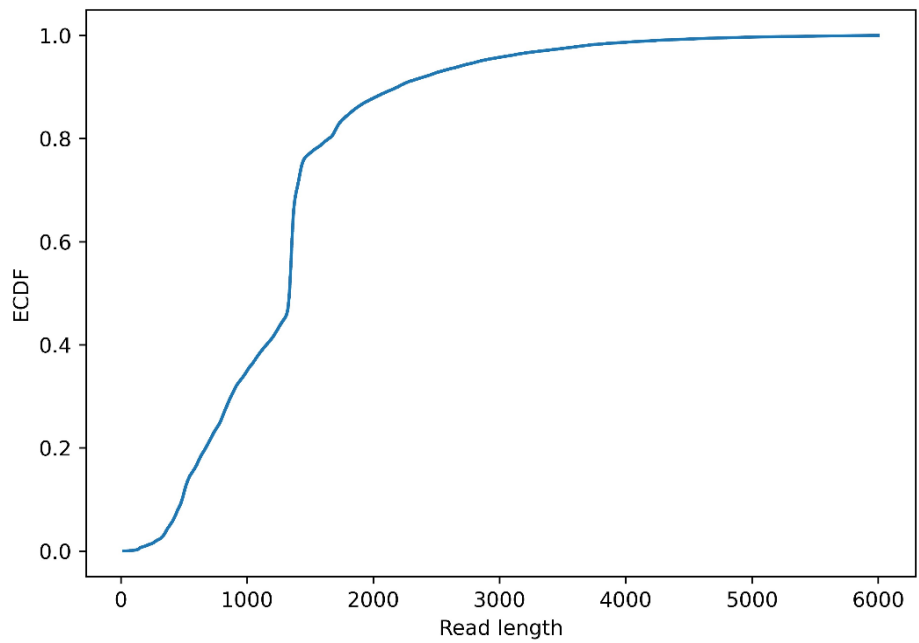

**Fig. S2.** Visualization of the max distance (A) and piecewise linear regression (B).

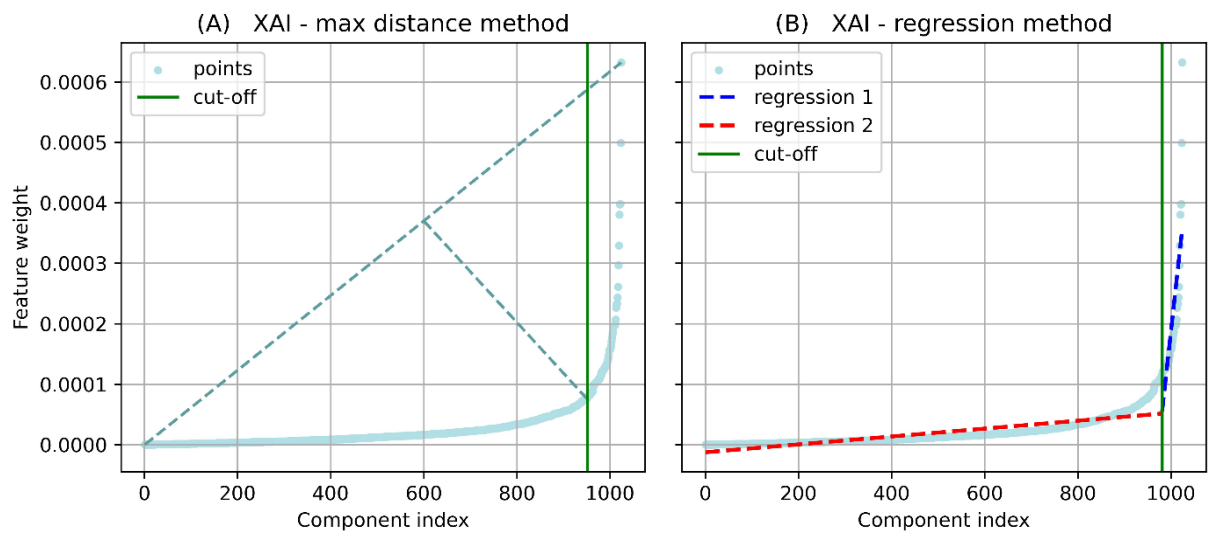

*Methods for determining the cut-off point, using the XAI ranking as an example.*

**Fig. S3.** Compatibility of the feature selection methods for ranking dictionary components of the highest importance.

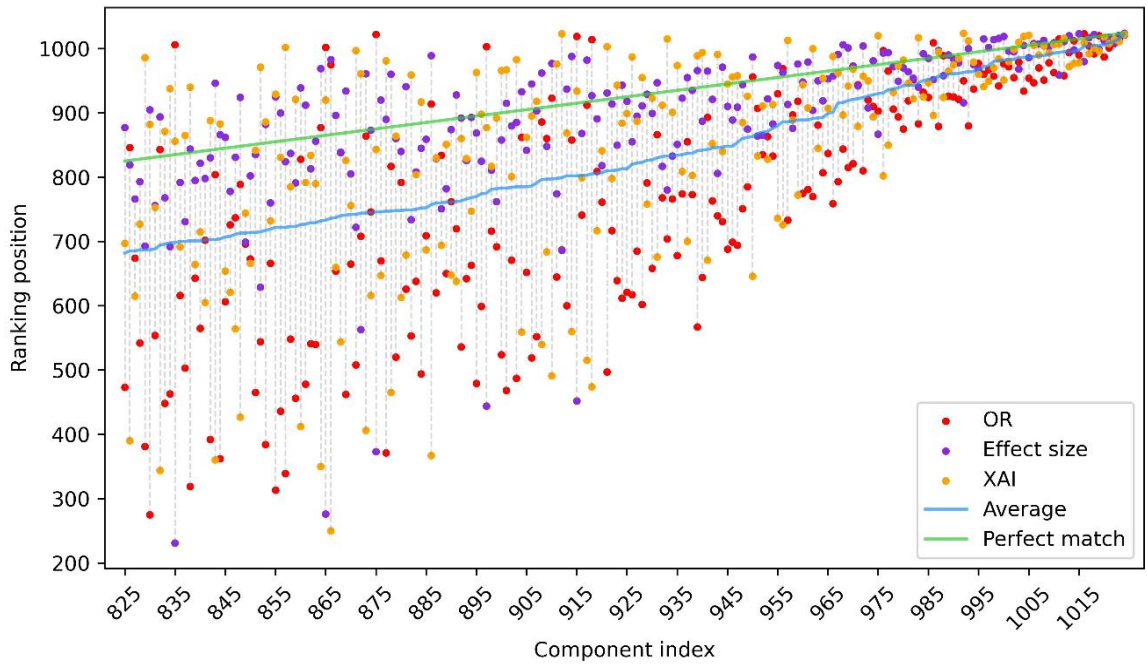

*Results are presented for the first LOSOCV repetition. Ranks from 825 and above.*

**Fig. S4.** Methods' compatibility for selecting the dictionary components with top 30 weights.

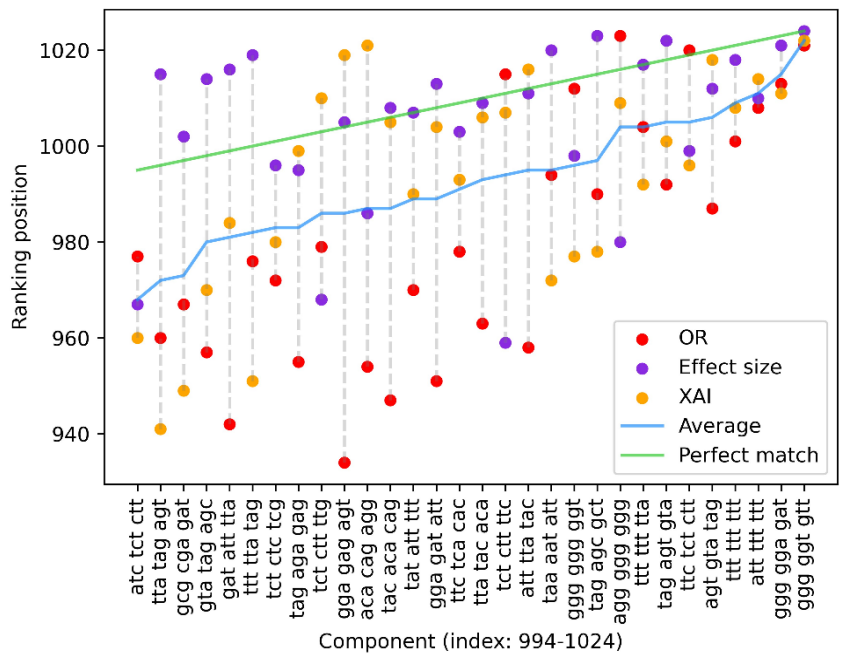

*Results are presented for the first LOSOCV repetition.*

**Fig. S5.** Consistency in selecting the most important dictionary components across all ranking methods.

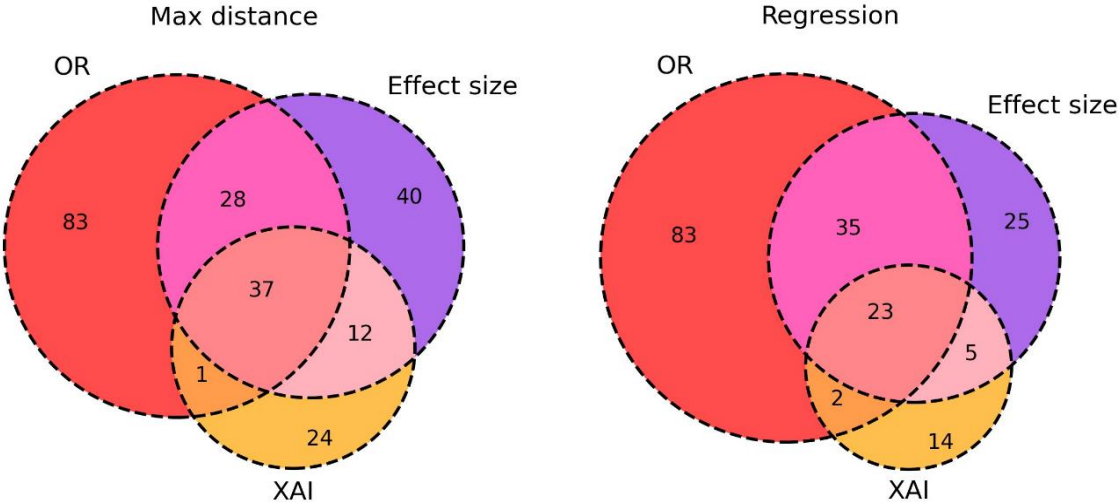

*Diagrams for the first LOSOCV repetition.*